\RequirePackage{fix-cm}
\documentclass[smallextended]{svjour3}       
\smartqed  
\usepackage{graphicx}
\begin{document}

\title{A Simple Voting Protocol on Quantum Blockchain
}


\author{Xin Sun        \and
        Quanlong Wang 
        \and Piotr Kulicki
}


\institute{Xin Sun \at
              Department of the Foundations of Computer Science, \\
The John Paul II Catholic University of Lublin, Lublin, Poland\\
              \email{xin.sun.logic@gmail.com}           
           \and
           Quanlong Wang \at
            Department of Computer Science, \\ University of Oxford, Oxford, UK          
           \and
           Piotr Kulicki \at
              Department of the Foundations of Computer Science, \\
The John Paul II Catholic University of Lublin, Lublin, Poland
}

\date{Received: date / Accepted: date}

\maketitle

\begin{abstract}

This paper proposes a simple voting protocol based on quantum blockchain. Despite being simple, our voting protocol is anonymous, binding, non-reusable, verifiable, eligible, fair and self-tallying. Our protocol is also realizable by the current technology.

\end{abstract}

\section{Introduction}
Many voting protocols based on classical cryptography have been developed and successfully
applied since Chaum \textit{et a}l \cite{Chaum88}. However, the security of classical cryptography is based on the unproven complexity of some computing tasks, such as factoring large numbers. The research in quantum computation shows that quantum computers are able to factor large numbers in a short time, which
means that classical cryptography is already insecure. To react to the threat of quantum computers, a number of quantum voting protocols have been developed in the last decade \cite{Hillery06,Vaccaro07,Li08,Horoshko11,Li12,Jiang12,Tian16,Wang16,Bao17,Thapliyal7}.

To be reliable and useful in practice, voting protocols should satisfy some
desirable properties such as: 
\begin{enumerate}

\item Anonymity. Only the voter knows how he or she votes.

\item Binding. Nobody can change a ballot after its submission.

\item Non-reusability. Every voter can vote only once.

\item Verifiability. Every voter can verify whether his or her ballot has been counted properly.

\item Eligibility. Only eligible voters can vote.

\item Fairness. Nobody can obtain a partial ballot tally before the stage of tally.

\item Self-tallying. Everyone who is interested in the voting result can tally ballots by himself or herself.

\end{enumerate}

To the best of our knowledge, among all the existing quantum voting protocols, only Wang \textit{et al.} \cite{Wang16} satisfy all the above requirements. 
%
%
However, their protocol is difficult to be implemented with the current technology. In this paper, our aim is to develop a voting protocol which satisfies all the above properties, in addition to be implementable by the current technology.

An important feature of our protocol is that it utilizes quantum-enhanced logic-based blockchain developed in \cite{Sun18Blockchain}. 
It turns out that blockchain can significantly simplify the design of protocols of electronic voting. 
A quantum bit commitment protocol is also needed to ensure some desirable properties of voting. We choose 
a cheat-sensitive quantum bit commitment protocol from \cite{SunWang18} for this purpose because it is efficient and implementable by the current technology.

We review some background knowledge on quantum blockchain and quantum bit commitment in Section \ref{Background}. Then,  in Section \ref{Voting on quantum blockchain}, we proposed our voting protocol. We conclude this paper in Section \ref{Conclusion and future work}.


\section{Background}\label{Background}

\subsection{Quantum blockchain}

A blockchain is a distributed, transparent and append-only database which enables achieving consensus in a large decentralized network of agents who do not trust each other. It is distributed in the sense that every miner (an agent in charge of updating the database) has an identical copy of the database. One of the most prominent applications of blockchain technology is for cryptocurrencies, such as Bitcoin \cite{Nakamoto08}. Another important application is the implementation of self-executable smart contracts  \cite{Szabo97}.

In the quantum blockchain system presented in \cite{Sun18Blockchain}, which we are going to use for our voting protocol, it is assumed that each pair of nodes (agents) is connected by an authenticated quantum channel and a not necessary authenticated classical channel.  
Each pair 
can establish a sequence of secret keys by using quantum key distribution \cite{Bennett84}. Those keys will later be used for message authentication.
%

Updates (new transactions/new messages) of a blockchain are initiated by those nodes who wish to append some new data to the blockchain. The classical data of an update is sent via classical channels to all miners, while the quantum data of the update is sent via quantum channels. Each miner checks the consistency of the update  with respect to their local copy of the database 
and forms an opinion regarding the update's admissibility. 
Then all the miners apply the quantum honest-success Byzantine agreement protocol 
introduced in  \cite{Sun18Blockchain} to the update, arriving at a consensus regarding the correct version of the update and whether the update is admissible. Finally, the update is added to the copies of the database of every node if at least  half of the miners agree that the update is admissible.

\subsection{Quantum bit commitment}

Bit commitment, used in a wide range of cryptographic protocols (e.g. zero-knowledge proof, multiparty secure computation, and oblivious transfer), typically consists of two phases, namely: commitment and opening.
In the commitment phase, Alice the sender chooses a bit $a$ ($a = 0$ or $1$) which she wishes to commit to the receiver Bob. Then Alice presents Bob some evidence about the bit. The committed bit cannot be known by Bob prior to the opening phase. Later, in the opening phase, Alice discloses 
some information needed for reconstructing $a$. Bob then reconstructs a bit $a'$ using Alice's evidence and announcement. A correct bit commitment protocol will ensure that $a' = a$. A bit commitment protocol is concealing if Bob cannot know the bit Alice committed before the opening phase and it is binding if Alice cannot change the bit she committed after the commitment phase. 

The first quantum bit commitment (QBC) protocol was proposed in 1984 by Bennett and Brassard  \cite{Bennett84}. A QBC protocol is unconditionally secure if any cheating can be detected with a probability arbitrarily close to 1. Here, Alice's cheating means that Alice changes the committed bit after the commitment phase, while Bob's cheating means that Bob learns the committed bit before the opening phase. A number of QBC protocols are designed to achieve unconditional security, such as those of \cite{Brassard90,Brassard93}. However, according to the Mayers-Lo-Chau (MLC) no-go theorem \cite{Mayers97,LoChau97}, unconditionally secure QBC  in principle can never be achieved.

Although unconditional secure QBC is impossible, several QBC protocols satisfy some other notions of security, such as cheat-sensitive quantum bit commitment (CSQBC) protocols \cite{Hardy04,Buhrman08,Shimizu11,Li14,ZhouSun18} and relativistic QBC protocols \cite{Kent11,Adlam15}. In CSQBC protocols, the probability of detecting cheating is merely required to be non-zero. With this less stringent security requirement, many  QBC protocols which are not unconditional secure are regarded as secure in the cheat-sensitive sense and they are also resilient to the attack of quantum computers.

In Sun and Wang \cite{SunWang18} a CSQBC is proposed which is more secure and efficient than all other existing CSQBC protocols.
 According to Tatar \textit{et al} \cite{Tatar18},  this protocol is also piratically resilient to the entanglement attack, which damages the unconditional security of many QBC protocols \cite{Mayers97,LoChau97}. Moreover, this protocol is implementable by the current technology. Although it is discovered that all CSQBC protocols are flawed if there is no mechanism of punishment, this flaw is fixed by running CSQBC on quantum blockchain, as it is shown in \cite{Sun18Blockchain}.  


\section{Voting on quantum blockchain} \label{Voting on quantum blockchain}

In the simplest setting of voting, $n$ voters vote on an issue. Every voter $V_i$ has a private binary value $v_i\in \{0,1\}$, where  $v_i = 0$ means disagree and $v_i =1$ means agree. The following is our protocol for simple voting, of which the structure is similar to (and simpler than) the voting protocol on the Bitcoin blockchain \cite{Zhao15,Tian17}. It consists of two phases: the ballot commitment phase and the tally phase.

\begin{enumerate}
\item Ballot Commitment.  

\begin{enumerate}
\item For each $i\in \{1,\ldots, n\}$, voter $V_i$ generates the $i$-th row of an $n \times n$ matrix of integers $r_{i,1},\ldots r_{i,n}$, whose sum $\sum_j r_{i,j}$ and 0 are congruent modulo $n+1$. That is,   $\sum_j r_{i,j} \equiv 0  \mbox{ } ( mod \mbox{ } n+1)$.

\item For each $i$ and $j$, voter $V_i$ sends $r_{i,j}$ to $V_j$ via quantum secure communication \cite{Bennett84,Zhou18}.

\item Now for each $i$, voter $V_i$ knows the $i$-th column $r_{1,i},\ldots, r_{n,i}$. Then  he computes his masked ballot $\widehat{v_i} \equiv v_i + \sum_j r_{j,i}  \mbox{ } ( mod \mbox{ } n+1) $. $V_i$ commits $\widehat{v_i}$ to every miner of the blockchain by the CSQBC protocol of Sun and Wang \cite{SunWang18}.

\end{enumerate}

\item Tally by decommitment.

\begin{enumerate}
\item For each $i$, $V_i$ reveal $\widehat{v_i}$ to every miner of the blockchain by opening his commitment.

\item All the miners run the quantum honest-success Byzantine agreement protocol \cite{Sun18Blockchain} to achieve a consensus of on the masked ballot $\widehat{v_1}, \ldots , \widehat{v_n}$.

\item The result of voting is obtained by calculating $ \sum_i \widehat{v_i}$, which equals to $ \sum_i  v_i $ because $ \sum_i \widehat{v_i} \equiv \sum_i ( v_i + \sum_j r_{j,i} ) \equiv \sum_i  v_i + \sum_{i,j} r_{j,i}  \equiv \sum_i ( v_i + \sum_j r_{i,j} ) \equiv  \sum_i  v_i  \mbox{ } ( mod \mbox{ } n+1)$.

\end{enumerate}

\end{enumerate}

\subsection{Security analysis}

Our voting protocol satisfies the following security requirements: 
\begin{enumerate}

\item Anonymity. 

This is because quantum secure communication prohibits other voters to gain any knowledge about a voter's ballot.

\item Binding.

Other voters cannot change a voter's ballot because of the authentication procedure of the blockchain. The voter himself cannot change his submitted ballot because of the binding property of quantum bid commitment.

\item Non-reusability. 

This is because the blockchain technology  solves the double spending problem: no voter can successfully uploaded two different ballots to the blockchain.

\item Verifiability. 

Every voter can check if his masked ballot is successfully upload to the blockchain because the blockchain is a transparent database.

\item Eligibility. 

This is ensured by the authentication procedure of the blockchain: only authenticated voters can successfully communicate to the miners.

\item Fairness. 

This is ensured by the concealing property of quantum bit commitment.

\item Self-tallying. 

This is because the blockchain is a transparent database. All data on the blockchain is accessible to every interested user.

\end{enumerate}

\section{Conclusion and future work}\label{Conclusion and future work}

This paper proposes a simple voting protocol based on quantum blockchain. Besides being simple, our voting protocol is anonymous, binding, non-reusable, verifiable, eligible, fair and self-tallying. Our voting protocol is also realizable by the current technology.

We have demonstrated that quantum blockchain can significantly simplify the task of electronic voting. In the future, we are interested in applying quantum blockchain to other fields such as auction and mechanism design. We believe quantum blockchain will also simplify these tasks.



\bibliographystyle{spmpsci}      
\bibliography{reference}   

%
%

\end{document}